\documentclass[%
 reprint,
 amsmath,amssymb,
 aps,
]{revtex4-1}

\usepackage{graphicx}
\usepackage{dcolumn}
\usepackage{bm}

\begin{document}

\title{Non collinear Magnetism and Phonon Dispersion Relation in Vacancy Induced Phosphorene Monolayer}
\author{Sushant Kumar Behera and Pritam Deb}
 \email{Corresponding Author: pdeb@tezu.ernet.in}
 \affiliation{Advanced Functional Material Laboratory, Department of Physics, Tezpur University (Central University), Tezpur-784028, India.}

\date{\today}

\begin{abstract}
We have studied the electronic, magnetic and linear phonon dispersion behavior of Phosphorene monolayer using ﬁrst principle based ab initio method. Phosphorene monolayer is a semiconducting system with a dimensional 
dependent variable range of band gap. Vacancy has been done to study the geometry and physical behavior of the monolayer system. Pristine, vacancy induced monolayer and vacancy induced doped monolayer are included in the 
calculation. Dopant concentration has been well checked via optimization algorithm to maintain the dilute magnetic semiconducting behavior of the monolayer system. Density of states and partial density of state indicates 
the contribution of individual orbitals in the system. Band closing nature in observed in vacancy and doped vacancy states indicating closed dense states and metallic behavior of the perturbed phases. Both antiferromagnetic 
and ferromagnetic ordering is included in our calculation to get a charm of both ordering in the physical properties of the system. Landau energy level distribution is mapped via Fermi surface with linear dispersion relation 
in terms of phonon vibrational density of states and linear dispersion relations. The results of linear phonon density of states corroborating with electronic density of states. 
\begin{description}
\item[keywords]
Phosphorene, Phonon Dispersion relation, Non-collinear Magnetism, Landau Energy Level, Mapping
\end{description}
\end{abstract}
\maketitle
\section{Introduction}
Two dimensional (2D) atomic crystals are stable materials carrying many different interesting properties unlike their 3D counterparts with potential applications\cite{1,2}. Electronic properties in 2D materials vary from semiconducting 
to metallic depending on the composition and thickness. Among the 2D materials, Phosphorene, the most stable allotrope, is a single or few layer of black phosphorus with a stacked puckered structure where individual layers held 
together by van der Waals forces. Structural anisotropy is supported via orthogonal unit cell with armchair and zigzag configurations along the x and y direction, respectively\cite{3}. As a result, it possess a unique structural 
characteristics along the armchair direction and a bilayer configuration along the zigzag direction. This unique structural arrangement is the key factor of the novel anisotropic physical (i.e. electronic structure, magnetic 
moment development, phonon dispersion, etc.) properties of phosphorene monolayer. Vacancy and doping in pristine monolayer help in modulating the physical phenomena of the pristine sheet. As a result, the overall electronic, 
magnetic and mapping of Landau energy levels in terms of Fermi energy are modulated for enhanced characteristics.\\

\begin{figure*}
\includegraphics[width=10cm,height=6cm]{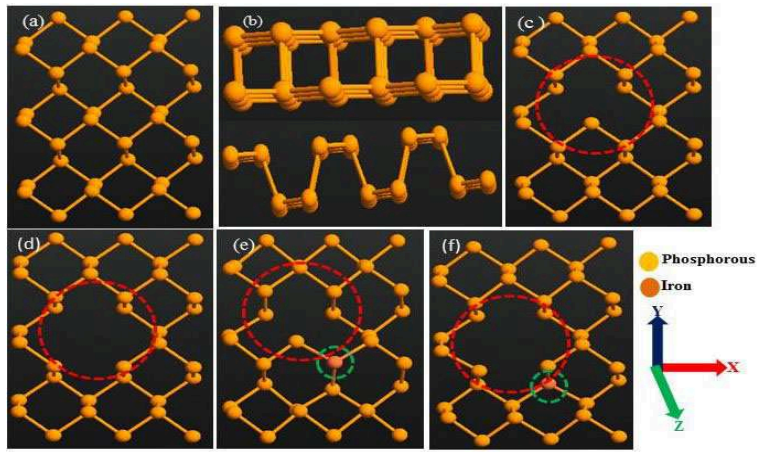}
\caption{\label{Fig:wide}The optimized geometry of phosphorene monolayer in (a) pristine phosphorene, (b) one in the top is the zigzag direction of pure phosphorene and the bottom one is the armchair direction of pure phosphorene 
with 36 atoms, (c) phosphorene with monovacancy, (d) phosphorene with divacancy having a total of 34 atoms, (e) doped monovacancy phosphorene and (f) doped divacancy phosphorene. The red circle denotes the vacancy created and the 
green circle denotes the dopant i.e. the Fe atom.}
\end{figure*} 

Diluted magnetic semiconductors (DMS) are compounds with semiconducting nature where magnetic atoms are used to replace their constituent atoms\cite{4}. It is our current interest to gain moment at room temperature in a few classes of 
layered DMS materials. Still, induced magnetism in 2D semiconducting materials has remained an interesting and unexplored field of current research. \\

Recently, first-principles based theoretical calculations are followed to understand the magnetic phenomena and phonon dispersion frequency of layered DMS materials. Moreover, low symmetry and high anisotropy of phosphorous atoms 
show interesting platform to study magnetic behavior and their phonon frequency in a form of puckered honeycomb lattice in phosphorene monolayer. These reports stimulate lots of research efforts in evaluating the magnetic moment 
development and phonon dispersion relation in monolayer phosphorene via doping or vacancy for potential application purposes. In this present work, we use density functional theory combined with PHONON codes to calculate electronic, 
magnetic and phonon dispersion properties. Moreover, we study the electronic structure, magnetic behavior and phonon dispersion in phosphorene monolayer with vacancy and doping iron atom.\\

\begin{figure*}
\includegraphics[width=12cm,height=7cm]{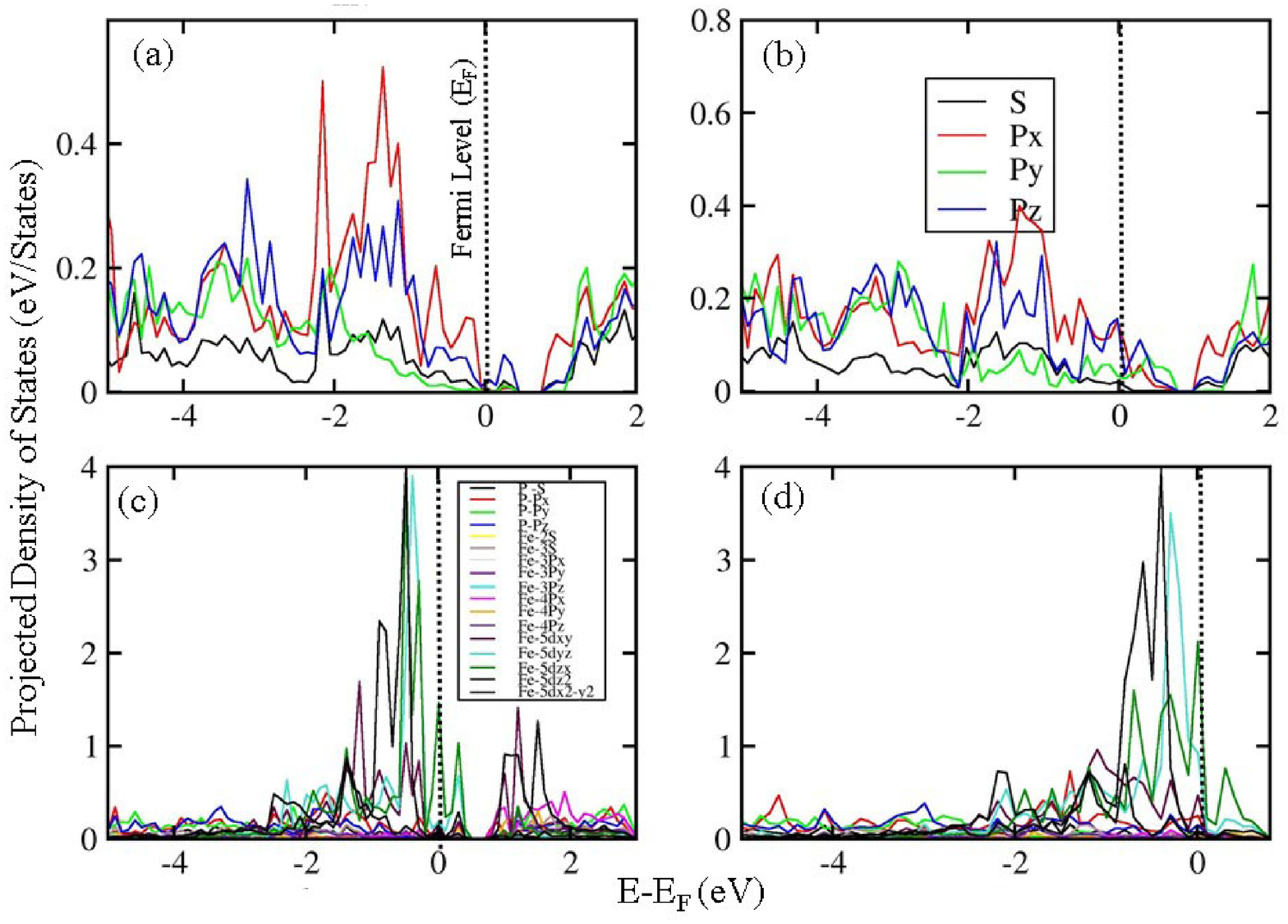}
\caption{\label{Fig:wide}Projected density of states of (a) MV, (b) DV, (c) DMV and (d) DDV. The dotted black line presents the Fermi energy ($E_F$) level.}
\end{figure*} 

\section{Methodology}
The density functional theory calculations were done by plane-wave based Quantum Espresso (QE) code with PAW potential and PBE-GGA exchange correlations\cite{5}. We maintain 20 $\AA{}$ of vacuum to minimize closest atomic interactions in the 
monolayer in supercell form. We use 48 eV as cutoff energy for geometry optimization and $9\times9\times1$ grid for sampling the Brillouin zone with Monkhorst-Pack scheme in a $3\times3\times1$ supercell. For the projection of the 
density of states (PDOS), we use a $27\times27\times1$ grid for Brillouin-zone integrations. The band structure is plotted along the high symmetric points of $\Gamma$, M, K and $\Gamma$ in reciprocal space. Then, we could get the phonon dispersion 
the systems using the PHONON package integrated in the Quantum Espresso package. We take five systems during our simulation, such as pristine phosphorene monolayer (Pristine), monovacancy in monolayer (MV), divacancy in monolayer (DV), 
iron doped in MV named as doped monovacancy in monolayer (DMV) and iron doped in DV named as doped divacancy in monolayer (DDV). The five systems are simulated under three different magnetic orientations, e.g. nonmagnetic (NM), 
antiferromagnetic (AFM) and ferromagnetic (FM) ordering.\\

\section{Results and Discussions}
The geometry optimized structures of the phosphorene monolayer (shown in Figure 1) are obtained using geometry optimization with the quasi-Newtonian BFGS algorithm. The mapping of Landau energy level is found for lowest iteration steps
to gain stability of the surface atomic configuration. The constituent atoms are mobile during calculation to fix the minimum energy based stable configurations. \\

\begin{figure}
\includegraphics[width=8.5cm,height=4cm]{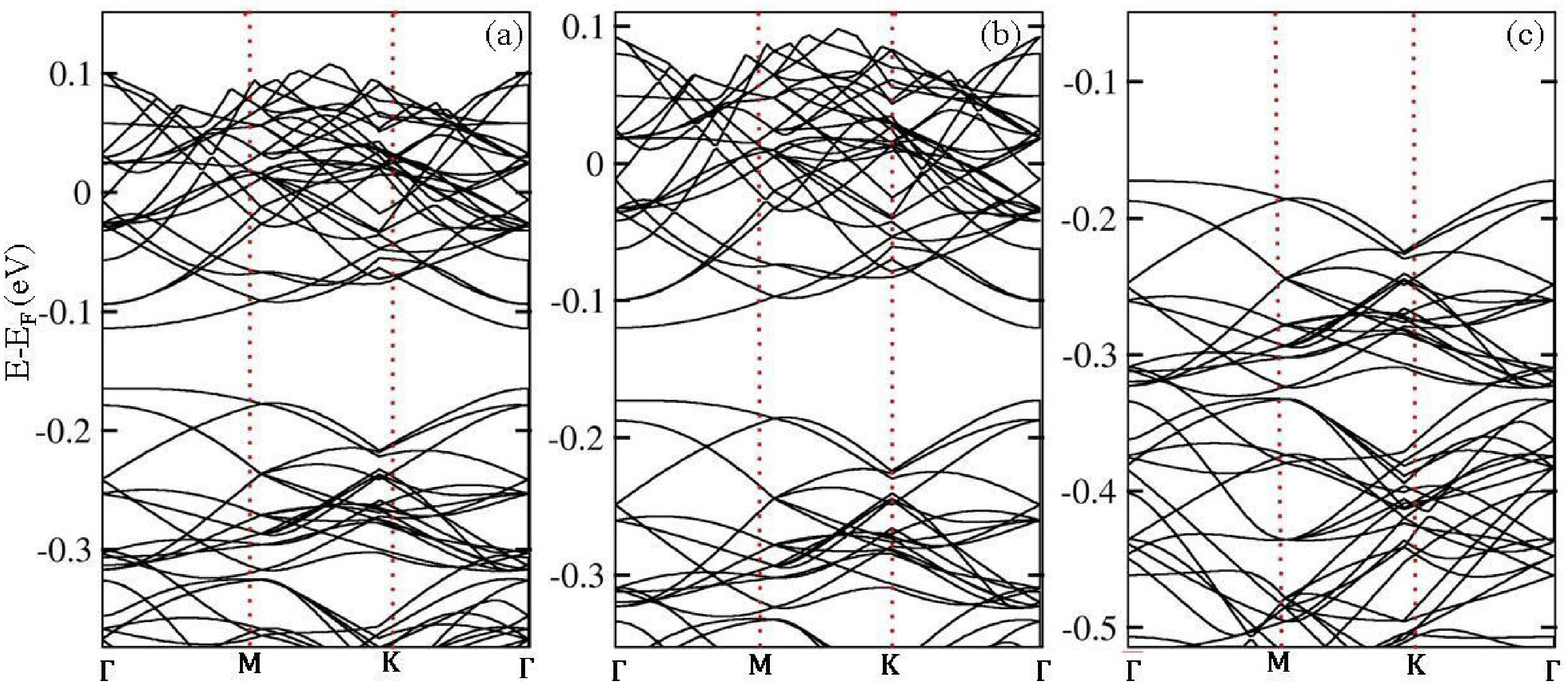}
\caption{\label{Fig:wide}Band structure of pristine Phosphorene in different orientations of (a) NM, (b) AFM and (c) FM. The dotted vertical lines presents the symmetric points in reciprocal space. The high symmetric points are 
$\Gamma$-M-K-$\Gamma$ in the calculation. The dotted red line presents the Fermi energy ($E_F$) level.}
\end{figure} 

Atomic orbitals are projected (Figure 2) to suitable energy level to identify near Fermi level activity of the vacancy induced and doped vacancy based monolayer systems. Transition from semiconducting to metallic phase is found from 
the bandgap closing behaviour (Figure 3) of the systems apart from pristine one.   \\

The defect states develops finite nonzero magnetic moments due to imbalanced surface spins in the pristine monolayer. Doped Fe atom enhances the moment value showing suitability towards room temperature dilute ferromagnetic 
semiconductor. The phonon frequency gap (94 c$m^{-1}$ for pristine, 90 c$m^{-1}$ for MV and 66.7 c$m^{-1}$ for DV) goes on decreasing as we go on creating vacancy in the pristine phosphorene to obtain ﬁrst monovacancy and then divacancy.\\

\section{Conclusions}
In summary, we sum up the results of electronic and magnetic properties of the systems under various magnetic ordering. Overlapping projection states of the atomic orbitals near Fermi level proves the evidence of metallic nature 
in the monolayer systems. Moreover, the bandgap closing behavior of the vacancy induced and doped systems compared to pristine system indicates the phase transformation from semiconducting to metallic phase which is suitable for 
magnetic property study. Phonon frequency gap and lattice dynamics in terms of Landau energy mapping can be successfully tuned by creating defects in the system in the form of vacancy and doping. Thus phosphorene has a bright 
future as a room temperature DMS layered material and Phononic crystal in radio-frequency communications and ultrasound imaging for medicine and nondestructive testing. \\

\begin{acknowledgments}
SKB acknowledges DST, Govt of India for providing INSPIRE Fellowship. The authors would like to thank Tezpur University for providing HPCC facility.
\end{acknowledgments}

\nocite{*}

\bibliography{manuscript}

\providecommand{\noopsort}[1]{}\providecommand{\singleletter}[1]{#1}%
\begin{thebibliography}{5}%
\makeatletter
\providecommand \@ifxundefined [1]{%
 \@ifx{#1\undefined}
}%
\providecommand \@ifnum [1]{%
 \ifnum #1\expandafter \@firstoftwo
 \else \expandafter \@secondoftwo
 \fi
}%
\providecommand \@ifx [1]{%
 \ifx #1\expandafter \@firstoftwo
 \else \expandafter \@secondoftwo
 \fi
}%
\providecommand \natexlab [1]{#1}%
\providecommand \enquote  [1]{``#1''}%
\providecommand \bibnamefont  [1]{#1}%
\providecommand \bibfnamefont [1]{#1}%
\providecommand \citenamefont [1]{#1}%
\providecommand \href@noop [0]{\@secondoftwo}%
\providecommand \href [0]{\begingroup \@sanitize@url \@href}%
\providecommand \@href[1]{\@@startlink{#1}\@@href}%
\providecommand \@@href[1]{\endgroup#1\@@endlink}%
\providecommand \@sanitize@url [0]{\catcode `\\12\catcode `\$12\catcode
  `\&12\catcode `\#12\catcode `\^12\catcode `\_12\catcode `\%12\relax}%
\providecommand \@@startlink[1]{}%
\providecommand \@@endlink[0]{}%
\providecommand \url  [0]{\begingroup\@sanitize@url \@url }%
\providecommand \@url [1]{\endgroup\@href {#1}{\urlprefix }}%
\providecommand \urlprefix  [0]{URL }%
\providecommand \Eprint [0]{\href }%
\providecommand \doibase [0]{http://dx.doi.org/}%
\providecommand \selectlanguage [0]{\@gobble}%
\providecommand \bibinfo  [0]{\@secondoftwo}%
\providecommand \bibfield  [0]{\@secondoftwo}%
\providecommand \translation [1]{[#1]}%
\providecommand \BibitemOpen [0]{}%
\providecommand \bibitemStop [0]{}%
\providecommand \bibitemNoStop [0]{.\EOS\space}%
\providecommand \EOS [0]{\spacefactor3000\relax}%
\providecommand \BibitemShut  [1]{\csname bibitem#1\endcsname}%
\let\auto@bib@innerbib\@empty
\bibitem [{\citenamefont {Behera}\ and\ \citenamefont {Deb}(2017)}]{1}%
  \BibitemOpen
  \bibfield  {author} {\bibinfo {author} {\bibfnamefont {S.~K.}\ \bibnamefont
  {Behera}}\ and\ \bibinfo {author} {\bibfnamefont {P.}~\bibnamefont {Deb}},\
  }\href@noop {} {\bibfield  {journal} {\bibinfo  {journal} {RSC Advances}\
  }\textbf {\bibinfo {volume} {7}},\ \bibinfo {pages} {31393} (\bibinfo {year}
  {2017})}\BibitemShut {NoStop}%
\bibitem [{\citenamefont {S.~K.~Behera}\ and\ \citenamefont {Ghosh}(2017)}]{2}%
  \BibitemOpen
  \bibfield  {author} {\bibinfo {author} {\bibfnamefont {P.~D.}\ \bibnamefont
  {S.~K.~Behera}}\ and\ \bibinfo {author} {\bibfnamefont {A.}~\bibnamefont
  {Ghosh}},\ }\href@noop {} {\bibfield  {journal} {\bibinfo  {journal}
  {Chemistry Select}\ }\textbf {\bibinfo {volume} {2}},\ \bibinfo {pages}
  {3657} (\bibinfo {year} {2017})}\BibitemShut {NoStop}%
\bibitem [{\citenamefont {Hao}\ and\ \citenamefont {Chen}(2006)}]{3}%
  \BibitemOpen
  \bibfield  {author} {\bibinfo {author} {\bibfnamefont {F.}~\bibnamefont
  {Hao}}\ and\ \bibinfo {author} {\bibfnamefont {X.}~\bibnamefont {Chen}},\
  }\href@noop {} {\bibfield  {journal} {\bibinfo  {journal} {J. Appl. Phys.}\
  }\textbf {\bibinfo {volume} {2016}},\ \bibinfo {pages} {165104} (\bibinfo
  {year} {2006})}\BibitemShut {NoStop}%
\bibitem [{\citenamefont {et~al.}(2010)}]{4}%
  \BibitemOpen
  \bibfield  {author} {\bibinfo {author} {\bibfnamefont {K.~S.}\ \bibnamefont
  {et~al.}},\ }\href@noop {} {\bibfield  {journal} {\bibinfo  {journal} {Rev.
  Mod. Phys.}\ }\textbf {\bibinfo {volume} {82}},\ \bibinfo {pages} {1633}
  (\bibinfo {year} {2010})}\BibitemShut {NoStop}%
\bibitem [{\citenamefont {et~al.}(2009)}]{5}%
  \BibitemOpen
  \bibfield  {author} {\bibinfo {author} {\bibfnamefont {P.~G.}\ \bibnamefont
  {et~al.}},\ }\href@noop {} {\bibfield  {journal} {\bibinfo  {journal} {J.
  Phys. Condens. Matter.}\ }\textbf {\bibinfo {volume} {21}},\ \bibinfo {pages}
  {395502} (\bibinfo {year} {2009})}\BibitemShut {NoStop}%
\end{thebibliography}%

\end{document}